\documentclass[aps,prl,amsfonts,amssymb,amsmath,twocolumn,floatfix,A4]{revtex4}
\usepackage{graphicx}
\usepackage{amsmath}

\begin{document}
\title{Spectrum of Andreev Bound States in a Molecule Embedded Inside a Microwave-Excited Superconducting Junction}
\author{Jonas Sk\"oldberg, Tomas L\"ofwander, Vitaly S. Shumeiko, and Mikael Fogelstr\"om}
\address{Deptartment of Microtechnology and Nanoscience - MC2, Chalmers University of Technology,
S-412 96 G\"oteborg, Sweden.}
\date{Physical Review Letters {\bf 101}, 087002 (2008)}

\begin{abstract}
Non-dissipative Josephson current through nanoscale
superconducting constrictions is carried by spectroscopically sharp
energy states, so-called Andreev bound states. Although
theoretically predicted almost 40 years ago, no direct
spectroscopic evidence of these Andreev bound states exists to date.
We propose a novel type of spectroscopy based on embedding a superconducting
constriction, formed by a single-level molecule junction, in a microwave QED cavity environment.
In the electron-dressed cavity spectrum we find a
polariton excitation at twice the Andreev bound state energy, and a
superconducting-phase dependent ac Stark shift of the cavity
frequency. Dispersive measurement of this frequency shift can be
used for Andreev bound state spectroscopy. 
\end{abstract}

\maketitle

Supercurrents through mesoscopic or nanosized Josephson junctions
are mainly carried by spectroscopically sharp subgap bound states,
so-called Andreev bound states (ABS)\cite{andreev66}, as predicted
theoretically in Refs.~\cite{kulik69,ishii70}. When the weak link
is a quantum dot with a single or a few discrete levels 
supercurrents are also predicted to flow mainly
through ABS \cite{beenakker92}. ABS come in pairs, one state above
and one below the Fermi level. The two ABS of the pair have
opposite dispersion with the superconducting phase difference over
the junction and carry supercurrent in opposite directions across
the junction. These two states form a well-defined two-level system
that has been suggested as a qubit \cite{zazunov03,zazunov05}.
Before such a qubit can be realized, experimental detection and
characterization of this engineered two-level system should be
carried out. To our best knowledge, however, no one has to date
reported experiments with direct spectroscopic proof for the
existence of Josephson-current carrying ABS \cite{hess}.   

A single-wall carbon nanotube (swCNT) embedded between two
superconducting metal leads can support a supercurrent
\cite{kazumov99,buitelaar02,buitelaar03,jarillo06,jorgensen06,cleuziou06}. 
The electronic energy levels of the swCNT, formed through size
quantization, can be tuned in and out of resonance with the lead Fermi
levels by gating the swCNT. Such superconducting swCNT transistors
\cite{jarillo06} can further be switched from a Coulomb
blockade regime, to a Kondo regime, to a weakly interacting Fabry-Perot regime
by changing a back gate voltage \cite{jorgensen06}. The potential for
applications of such swCNT quantum-level junctions was demonstrated
through the fabrication and detailed functional control of a
nano-SQUID, involving two gated swCNT junctions with controlled on-off
states as well as controlled 0-$\pi$ SQUID states \cite{cleuziou06}.
Similar control has been demonstrated using semiconducting nanowire
junctions \cite{vanDam06,xiang06,marchenkov07}. 

\begin{figure}[b]
\includegraphics[width=0.95\columnwidth]{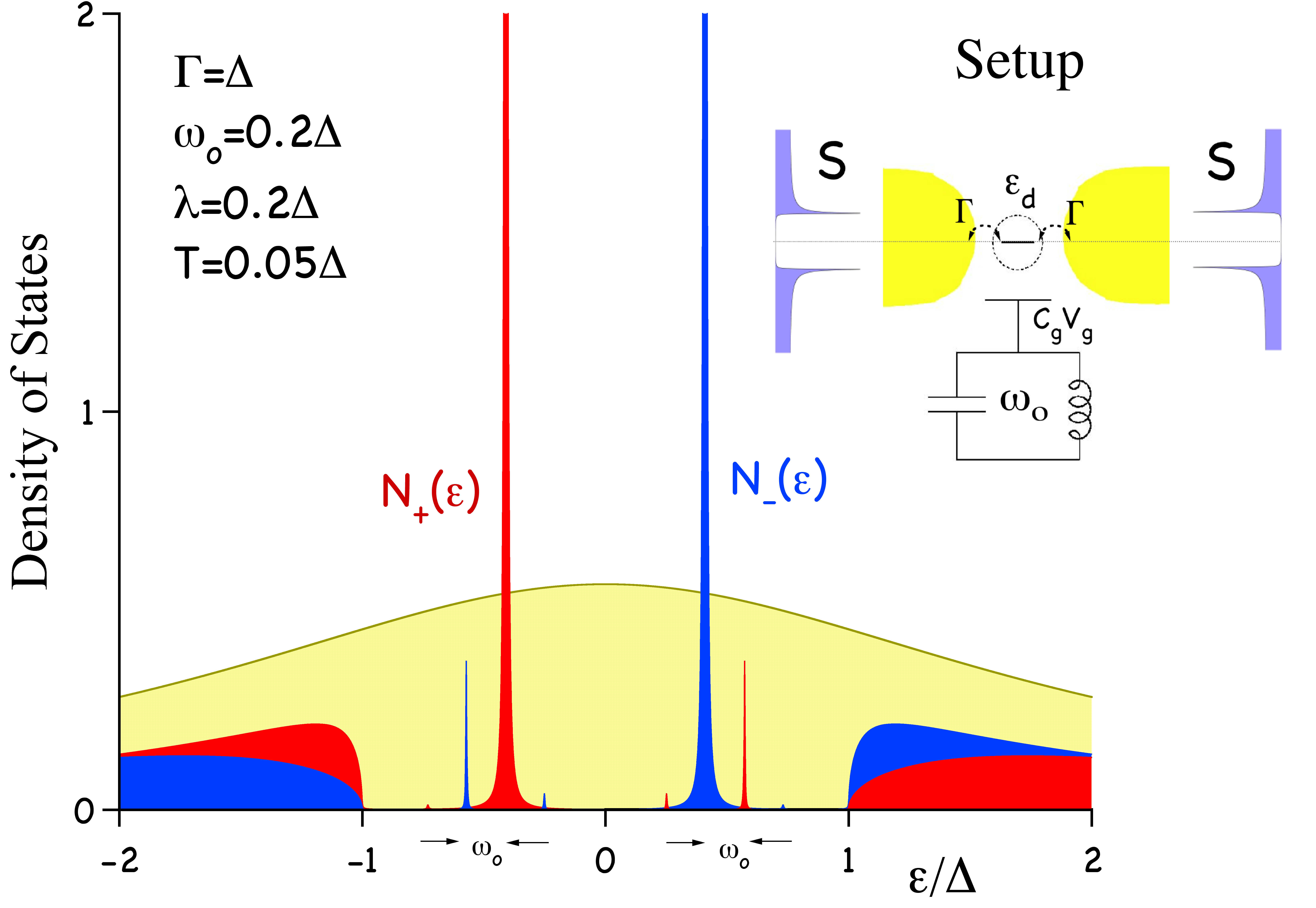}
\caption{
    Superconducting density of states for $\varphi=0.6 \pi$
    resolved into right, $N_+(\varepsilon)$, and left, $N_-(\varepsilon)$,
  current-carrying branches. The lightly shaded background is the 
  density of states of the resonant
  level when the contacts are in the normal state. The
  electron-oscillator coupling leads to dressed Andreev bound states
  with sidebands separated in energy by the oscillator frequency
  $\omega_o$. The inset shows a sketch of the considered system.}
\label{fig:Fig1}
\vspace*{-0.3truecm}
\end{figure}

We propose a method for direct ABS spectroscopy,
based on dispersive measurement of a polaritonic state formed by
the ABS strongly coupled to a QED cavity mode. Consider two
superconducting leads connected by a molecule with one resonant
level, as shown in Figure~1. This setup can be realized in the swCNT
transistor by tuning the voltage of the gate electrode
\cite{jarillo06,jorgensen06,cleuziou06}. The superconducting
proximity effect leads to a split of the resonant level into a pair
of ABS, where the ABS level splitting can be controlled by tuning
the superconducting phase difference in a SQUID setup. When the
gate electrode is coupled to an LC-oscillator, induced quantum
fluctuations of the gate potential leads to dressed Andreev bound
states. By tuning the ABS into resonance with the cavity, and
measuring the shift of the cavity base-frequency, the ABS energy
can be observed. The complete ABS energy dispersion with respect to
the superconducting phase can be extracted by using a cavity with a
variable base-frequency \cite{osborn07,castellanos07,palacios08,sandberg08,Wallquist06}. The suggested
dispersive method does not involve real interlevel transitions and
could be simpler to realize than methods based on microwave
absorption \cite{gorelik95}. Another advantage of the suggested
spectroscopy is that it makes it possible to directly access and
characterize individual ABS without measuring dc contact
current-voltage characteristics \cite{Saclay07}. 
Similar dispersive measurement method has earlier been applied to
detect quantum states of superconducting charge qubits
\cite{Wallraff04}.

The Hamiltonian for the coupled electron-oscillator system depicted in 
the inset in Figure \ref{fig:Fig1} is
\begin{equation}
\hat H = \hat H_{L} + \hat H_{R} + \hat H_{level} + \hat H_T + \hat H_{osc} + \hat H_{level-osc}
\label{systemH}
\end{equation}
and will be quantified in terms of creation operators for the three
excitations of the system: reservoir electrons,
$\hat c_{k\sigma,\alpha}^{\dagger}$ (reservoir $\alpha=L$, $R$, momentum
$k$, and spin $\sigma$); dot electrons, $\hat d_{\sigma}^{\dagger}$; and
the oscillator mode $\hat b^{\dagger}$. The two superconducting reservoirs
are described by a standard BCS Hamiltonian
$
\hat H_{\alpha}=\sum_{k\sigma} \xi_k \hat c_{k\sigma,\alpha}^{\dagger}\hat c_{k\sigma,\alpha} +
\sum_k \left( \Delta_{\alpha}\hat c_{k\uparrow,\alpha}^{\dagger}\hat c_{-k\downarrow,\alpha}^{\dagger}
+h.c.\right)
$,
where $h.c.$ denotes hermitian conjugate, $\xi_k$ is the quasiparticle
dispersion, and $\Delta_{\alpha}=\Delta(T){\rm e}^{i \varphi_{\alpha}}$ are
the order parameters of the reservoirs that have the same temperature
dependent gap $\Delta(T)$, but have a tunable superconducting phase
difference $\varphi=\varphi_{R}-\varphi_{L}$ between them.
The reservoirs are coupled by tunneling through a single
non-interacting molecular level described by
$
\hat H_{level} = \sum_{\sigma} \varepsilon_d \hat d_{\sigma}^{\dagger}\hat d_{\sigma},
$
where the level energy $\varepsilon_d(V_g)$ is tunable by the gate
voltage.
The lead-to-level tunneling is described by,
$
\hat H_T = \sum_{k\sigma,\alpha} \left(
v_{k\sigma,\alpha} \hat c_{k\sigma,\alpha}^{\dagger} \hat d_{\sigma} + h.c. \right)
$.
We consider symmetric, spin-independent coupling between the level
and the two leads, $v_{k\sigma,L}=v_{k\sigma,R}\equiv
v\delta_{k,k_F}$, which corresponds to an effective tunneling rate at
each barrier $\Gamma= \frac{\,\,|v|^2}{2 \pi} {\cal{N}}_F$, where
${\cal{N}}_F$ is the normal state density of states of the leads at
the Fermi-level. The oscillator mode with frequency $\omega_o$ is
described by the Hamiltonian
$
\hat H_{osc}=\omega_o \hat b^\dagger \hat b.
$
The coupling of the molecular level to the oscillator is described by
a linear interaction
$
\hat H_{level-osc}= \sum_\sigma \lambda (\hat b+ \hat b^\dagger) \hat d^\dagger_\sigma \hat d_\sigma,
$
where $\lambda$ is the coupling strength. We will assume throughout
this paper that the molecular level is aligned with the Fermi levels
of the leads, i.e. $\varepsilon_d=0$. 

We solve the Hamiltonian ($\ref{systemH}$) treating the tunneling to
infinite order in the hopping \cite{jauho94,cuevas96,cuevas01} 
and the electron-oscillator coupling perturbatively
in a self-consistent Born-approximation \cite{hyldgaard94,mitra04,viljas05,novotny05,zazunov06,zazunov06_2}.
The problem reduces to the coupled molecular level/oscillator system. 
Superconductivity modifies the electronic state on the molecular-level by splitting it
into two branches ($s=\pm1$) each described by a retarded Green's function
\begin{equation}
G^R_{s}(\varepsilon)=
\frac{\bar \Omega(\varepsilon)}
{z^R_s(\varepsilon) \varepsilon^R+s\Delta \cos(\varphi/2)},\quad
\label{retardedG}
\end{equation}
where the energy-renormalisation factor $z_s(\varepsilon)$ is defined by 
%
$
z^R_s(\varepsilon) \varepsilon=\tilde\varepsilon^R+
\bar\Omega(\varepsilon)\left(\varepsilon^R-\Sigma^R_{-s}(\varepsilon)\right),
$
%
with
$\bar\Omega(\varepsilon)=\sqrt{|\Delta|^2-(\tilde\varepsilon^R)^2}/ 2 \Gamma$.
The state of the oscillator is given by the retarded Green's function
\begin{equation}
D^R(\omega)=\frac{2 \omega_o}{(\tilde\omega^R)^2-\omega_o^2-2 \omega_o \Pi^R({\omega})}.
\quad
\label{retardedD}
\end{equation}
The electron-oscillator coupling enters via the retarded self-energies $\Sigma^R_{s}(\varepsilon)$
and $\Pi^R({\omega})$ which are determined self-consistently by numerical
iteration. $\Sigma^R_{s}(\varepsilon)$
and $\Pi^R(\omega)$ are functionals of $G^R_{s}(\varepsilon)$ and $D^R(\omega)$
[and of $G^K_s(\varepsilon)=-2 i {\rm Im}\, G^R_s(\varepsilon)\, n_e(\varepsilon)$, with $n_e(\varepsilon)=\tanh(\varepsilon/2T)$ 
and $D^K(\omega)=-2 i {\rm Im}\, D^R(\omega)\, n_b(\omega)$, with $n_b(\omega)= \coth(\omega/2T)$, $G^A_s=(G^R_s)^{*}$]
and defined as
\begin{eqnarray}
\!\!&\Sigma^R_{s}\!(\varepsilon)&\!\!\!=\!\!i \frac{\lambda^2}{2}\!
\!\big\lbrack
D^K\!(\omega)\!\circ\! G^R_{s}\!(\varepsilon-\omega)
\!+\! D^R\!(\omega)\!\circ\! G^K_{s}\!(\varepsilon-\omega)
\!\big\rbrack,
\label{electronsigma}\\
\!\!&\Pi^R\!(\omega)&\!\!\!=\!\!-i\lambda^2\!\!\sum_{s=\pm1}\!
\!\!\big\lbrack
G^R_{-s}\!(\varepsilon)\!\circ\! G^K_{s}\!(\varepsilon\!-\!\omega)\!+\!
G^K_{-s}\!(\varepsilon)\!\circ \!G^A_{s}\!(\varepsilon\!-\!\omega)
\!\big\rbrack.
\label{oscillatorpi}
\end{eqnarray}
where the convolution is defined as $a(x)\circ
b(x-y)=\int^\infty_{-\infty}\frac{d x}{2 \pi}a(x) b(x-y)$.
Upon reaching self-consistency the Josephson current is calculated as
\begin{equation}
I(\varphi)=\frac{e}{\hbar}\Delta \sin \frac{\varphi}{2}\,
\sum_{s=\pm 1} \int^{\infty}_{-\infty}\frac{d \varepsilon}{2\pi}
s\,{\rm{Im}}\left[
\frac{G^R_s(\varepsilon)}{\bar \Omega(\varepsilon)}\right] n_e(\varepsilon).
\label{josephsoncurrent}
\end{equation}

Inelastic coupling to the environment is introduced phenomenologically above
by $\tilde\epsilon^R=\epsilon^R+i\gamma$ and $\tilde\omega^R=\omega^R+i\kappa$.
The parameter $\gamma\ll\Delta$ describes a residual phase-breaking scattering rate
in the superconducting reservoirs. In absence of an electron-oscillator coupling it is
$\gamma$ that limits the life time of the ABS. $\kappa$ describes
the finite (long) life time of the mode originating from a finite (but
high) quality factor, $Q$, such that
$\kappa=\omega_0/Q\ll\omega_0$. In the present calculations we set
$\gamma=\kappa=10^{-3}\Delta$.
\begin{figure}[t]
\includegraphics[width=1.0\columnwidth]{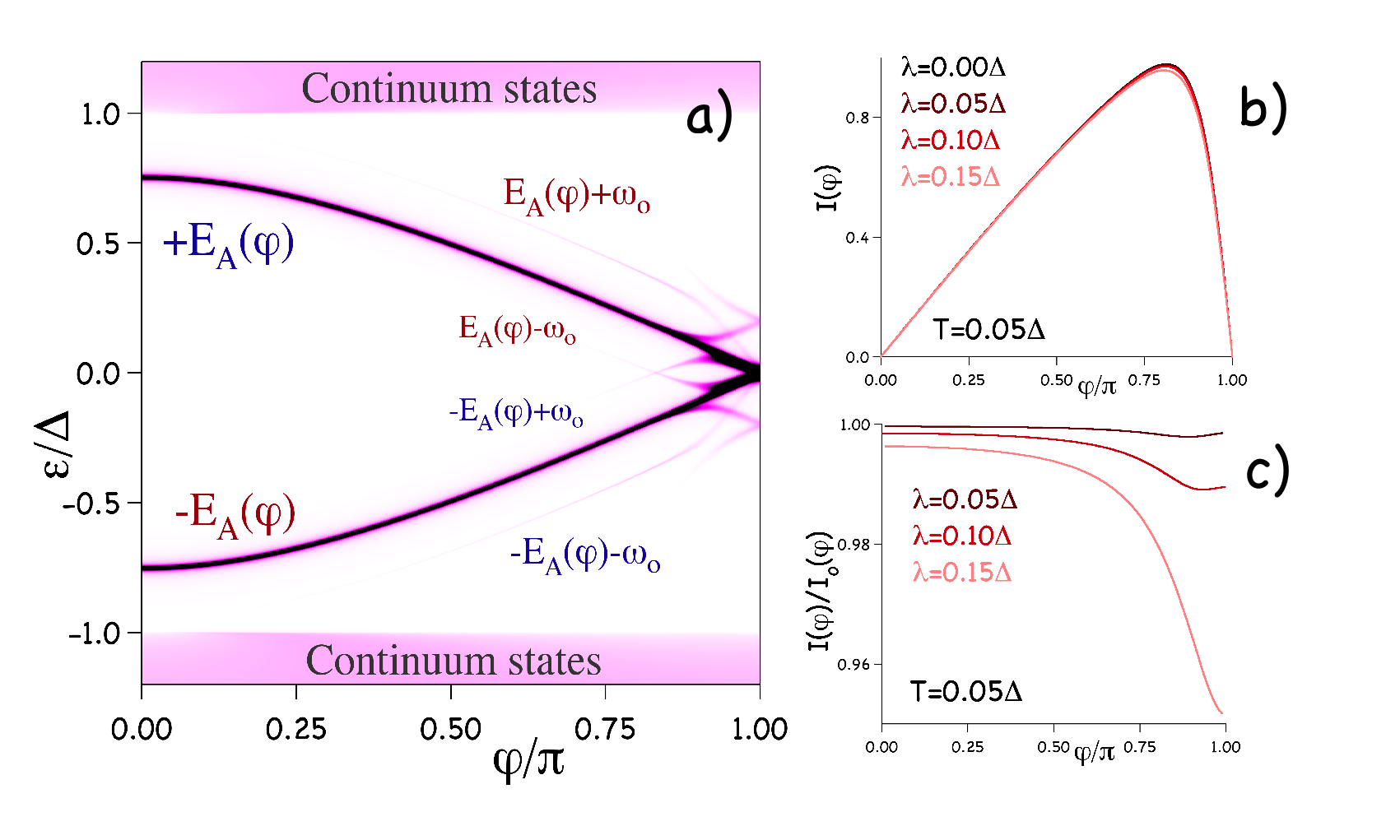}
\caption{
(a) The spectrum of Andreev-bound states as function of
superconducting phase-difference ($\varphi$) for
electron-oscillator coupling $\lambda=0.1\Delta$ and oscillator
frequency
  $\omega_o=0.2\Delta$. The tunneling rate is $\Gamma=\Delta$ and the
  temperature is $T=0.05\Delta$. The side bands to each ABS are due to
  the dressing of the ABS by the electron-oscillator coupling. The sidebands of the ABS
  with branch index $s=\pm1$ belong to the quaasiparticle branch with index $s\mp1$.
  (b) The current-phase relation for different
  electron-oscillator couplings for $T=0.05\Delta$. (c)
  The reduction of current due to coupling to the oscillator as a function of phase for 
  the same temperature and electron-oscillator couplings as in panel b).}
\label{fig:Fig2}
\vspace*{-0.3truecm}
\end{figure}

Without coupling to the oscillator, the Josephson effect in this
system is well known \cite{beenakker92}. %
The Green's function-amplitudes, $G^R_{s}$, describe two different
quasiparticle branches that form on the dot, one having an ABS
below ($s=+1$) and one having an ABS above ($s=-1$) the Fermi
level, see Figure~1. The two branches contribute to the Josephson 
current ($\ref{josephsoncurrent}$) 
in opposite manner. For sub-gap energies
$G^R_s(|\varepsilon| < \Delta)$ contributes to the current  
in the positive direction for $s=1$ (left to
right over the junction) and vice versa for the
$s=-1$ branch. The continuum part of 
$G^R_s(|\varepsilon| > \Delta)$ contributes to the
current in the opposite direction compared with its corresponding
sub-gap part. 

The electron-oscillator interaction couples the two quasiparticle branches.
This is seen explicitly in the energy-renormalization
factor $z_s$ of one branch which is modified
by the self-energy $\Sigma^R_{-s}$ of the other branch. 
In Figure~\ref{fig:Fig2} we show a fully self-consistent calculation of
the Andreev spectrum as function of superconducting
phase-difference. For the chosen parameters there is a resonance between the oscillator and the
ABS, i.e. $2E_A(\varphi) = \omega_o$, at $\varphi\approx0.9\pi$.
Away from resonance, the ABS is shifted
\begin{equation}
E_A\rightarrow \bar E_A =E_A+\frac{\lambda^2 {\cal A}_E^2}{2}
\bigg(\frac{\Phi_+}{2 E_A+\omega_o}+\frac{\Phi_-}{2 E_A-\omega_o}\bigg)
\label{ABSshift}
\end{equation}
as compared with the case without coupling to the oscillator, but with
retained phase-dependent spectral weight ${\cal A}_E$ of the state at $\varepsilon=\pm E_A(\varphi)$. 
The effective electron-oscillator coupling in the sub-gap region 
is given by the product $\lambda{\cal A}_E$.
In equation~($\ref{ABSshift}$) thermal occupation factors enter in the combinations
$\Phi_\pm=n_b(\omega_o)\pm n_e(E_A)$.
Apart from the shifted ABS we find satellite resonances
at $\varepsilon=s\bar E_A\pm\omega_o$ with spectral
weights
$\frac{1}{2}\lambda^2 {\cal A}_E^3\Phi_{\pm s}/(2sE_A\pm\omega_o)^2$.
It is only at resonance, $2E_A = \omega_o$, the satellite on either branch
with spectral weight $\propto  \Phi_-$ interfere with the main ABS of the same branch index.
This is seen as a precursor of an avoided crossing in Figure~\ref{fig:Fig2}. 

The Josephson current-phase relation is presented in panels b and c of 
Figure~\ref{fig:Fig2} for the case that the electron and oscillator
systems are in thermal equilibrium.  
Due to the different magnitudes
of the thermal factors $\Phi_\pm$ ($\Phi_+\approx 2$ and 
$\Phi_-\approx0\,\,$Êfor $T\lesssim E_A, \omega_o$), the current contribution of the
satellites is dominated by the satellite with weight $\propto
\Phi_+$. This satellite reduces the Josephson current by
$\sim\lambda^2 {\cal  A}_E^2/(2E_A+\omega_o)^2$, which is
of the order of a few percent of the full current for our parameter
values. There is no dramatic signature in
the current-phase relation of an emerging anticrossing at
resonance because interference occurs between states carrying
current in the same direction, and moreover, these states have
spectral weights shared between them drawn from the original ABS,
and the population of the states is largely phase-independent.
\begin{figure}[t]
\includegraphics[width=1.1\columnwidth]{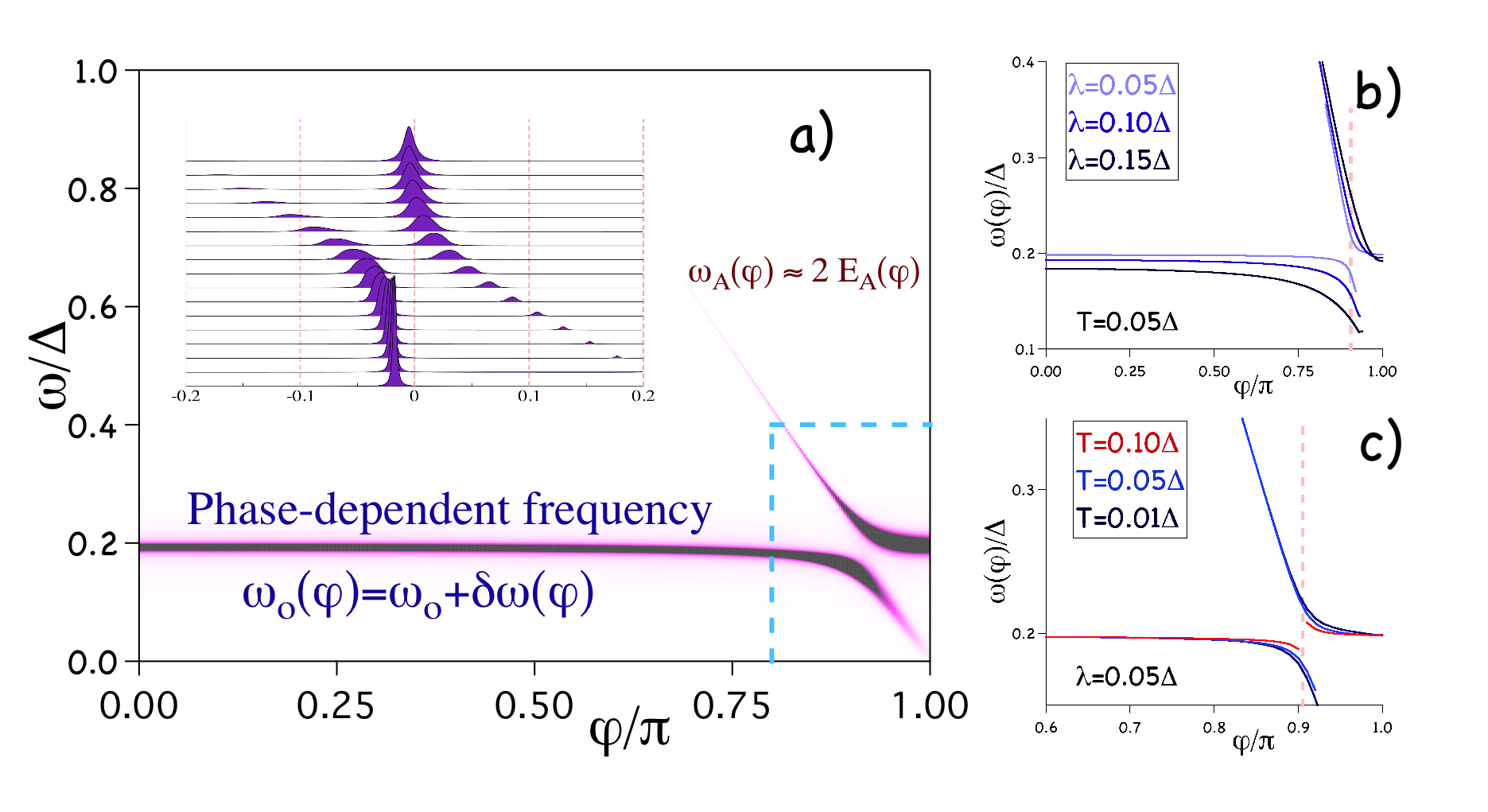}
\caption{
  (a) The spectra of the oscillator
  plotted as function of superconducting phase-difference with the
  same parameters as in Figure~\ref{fig:Fig2}.
  A mode emerges at $\omega_A(\varphi)=2E_A(\varphi)$ and develops 
  to a polariton when in resonance with the oscillator (inset, $0.8\pi\le\varphi\le\pi$ in steps of $0.0125\pi$). 
  The dashed box indicate where
  the spectra in the inset are taken.
  Away from resonance there is a Stark shift $\delta \omega(\varphi)$ of the base
  frequency. (b) The Stark shift is shown at different electron-oscillator
  coupling strengths for $T=0.05\Delta$ and (c) at
  different temperatures for a electron-oscillator coupling
  strength $\lambda=0.05\Delta$. In both cases
  $\omega_o=0.2\Delta$. The dashed line in panels b)-c) indicate the resonant phase.}
\label{fig:Fig3}
\vspace*{-0.3truecm}
\end{figure}

The Josephson current through the level modifies the oscillator
spectrum giving it a phase-dependence that is shown in
Figure~\ref{fig:Fig3}(a) for the same parameters as for the ABS
spectrum in Figure~\ref{fig:Fig2}. The polarization
$\Pi^R(\omega)$ gives the possible collective excitations that
are supported by the electronic system. Iterating the self-consistency equations once
we get
\begin{equation}
\Pi^R(\omega)= 2\lambda^2
\frac{{\cal A}_E^2}{{(\omega^R)}^2-4 {E_A}^2} n_e(E_A)
\label{oiphononpi}
\end{equation}
for the retarded phonon self-energy. As may be expected there is a
mode with the phase-dispersion $\omega_A=2E_A(\varphi)$ originating
from the transitions between the two ABS and the subsequent
emission or absorption of the energy $\omega_A$.
A spectral weight $\!\sim\! 2 \lambda^2 {\cal A}_E^2 \lbrack 4
\omega_o^2/(4 E_A^2-\omega^2)^2\rbrack n_e(E_A)$ for this
collective excitation is vanishingly small away from resonance.
Near resonance the oscillator mode and the excitation interact
strongly and an avoided crossing appears with a frequency split at
$2E_A=\omega_o$
\begin{equation}
\omega_o(\varphi)=\omega_o+\frac{\delta}{2}
\pm\sqrt{\left(\frac{\delta}{2}\right)^2+2\lambda^2 {\cal A}_E^2
n_e(E_A)} \label{omegaofphi}
\end{equation}
where $\delta=2E_A-\omega_o$ is the detuning. 
The analytic estimate of the split, $ 2 \lambda {\cal A}_E \sqrt{2
n_e(E_A)}\approx 0.096$, for a nearly phase-independent ${\cal
A}_E\sim 0.34$ when $\Gamma=\Delta$, is in good agreement with the
self-consistently determined split, $\approx 0.1 \Delta$, 
extracted from the inset in Figure~\ref{fig:Fig3}(a) at
$\varphi=0.9\,\pi$.
It is important to note that this split is significantly larger than
the intrinsic broadening of the spectral features ($\sim
\kappa$) and signals a strong-coupling regime in the sense of
cavity-QED, i.e. $\lambda {\cal A}_E\gg\kappa$.

Away from resonance, the electron-oscillator coupling gives a
negative phase-dependent Stark shift of the oscillator base
frequency. We plot this shift in Figure~\ref{fig:Fig3}(b)-(c) both
as a function of electron-oscillator coupling strength $\lambda$
for $T=0.05\Delta$ and as a function of temperature for
$\lambda=0.05\Delta$. Our analytic estimate in
equation~(\ref{omegaofphi}) gives in the limit of large detuning
\begin{equation}
\omega_o(\varphi)=\omega_o+\delta\omega(\varphi)
=\omega_o-2\frac{ \lambda^2 {\cal A}_E^2 n_e(E_A)}{2 E_A-\omega_o}
\end{equation}
This analytical expression for $\delta \omega(\varphi)$ is a good
approximation when the continuum contributions can be neglected,
i.e. for $\Gamma/\Delta\ll 1$, while in general one must use
numerics to extract $\delta \omega(\varphi)$. Dispersive
measurement of the acquired phase-dependent resonance frequency of
the oscillator, $\omega_o(\varphi)$, gives a possibility to detect
the position of the polariton resonance and hence define the energy of
the ABS. Furthermore, by sweeping the base frequency of the oscillator
the phase dispersion of the ABS energy can be detected.

Assume that we bridge two aluminium superconductors by a gated
swCNT, as shown in the inset of Figure~\ref{fig:Fig1}. The gate is
part of an LC-circuit which leads to an oscillation of the gate
voltage with frequency $\omega_o=1/\sqrt{LC}$. A practical value
for the resonator frequency is $\sim 10$ GHz, for which the
polariton resonance is located well inside the superconducting gap
(for aluminium $\Delta_{\rm Al}\approx$ 50 GHz), and at the
same time the ABS energy splitting is large compared to the
temperature below $100$mK. In our calculations we neglect
electron-electron interactions. This can be done if $\Gamma$ is
large compared to $\Delta$, which can experimentally be realized
by tuning $\Gamma$ by a back gate \cite{jorgensen06} to
approach the weakly interacting Fabry-Perot regime. The main
modification of the one-iteration approximation in the case $\Gamma
\gg \Delta$ is that ${\cal A}_E$ becomes strongly phase dependent,
tending to $(\Delta/2\Gamma) \sin(\varphi/2)$ as $\Gamma/\Delta$
grows. This gives the  effective coupling $\lambda {\cal A}_E\sim
\lambda\,\Delta /\Gamma$ close to the resonance. This coupling must
be large compared to the intrinsic oscillator damping,
$\lambda\Delta/\Gamma\gg\kappa=\omega_o/Q$, in order to resolve
$\delta \omega(\varphi)$.
The strength of the bare coupling $\lambda$ is determined by the
capacitive interaction between the gate and the dot, and it is
proportional to the ratio of corresponding capacitances,
$C_g/C_\Sigma$, and can be expressed through the oscillator
frequency as $\lambda =
(C_g/C_\Sigma)(E_C / 8E_L)^{1/4}\omega_o$ ($E_C$ and $E_L$ are
charging and inductive energies of the oscillator, respectively).
For swCNT superconducting contacts the gate capacitance can be
comparable with the capacitances of the contacts to the leads, having
values of tens of aF \cite{ Grove-Rasmussen07}. Thus the coupling
$\lambda$ can be on the order of $10$\% of the
oscillator frequency as assumed in our calculation. When the gate
of the contact is connected to a superconducting cavity, 
the quality factor of the oscillator can be of order 1000 - 10000 
\cite{osborn07,castellanos07,palacios08,sandberg08}. Given
our calculated value for effective ABS-oscillator coupling, $\lambda {\cal A}_E$, we get 
the ratio  $\lambda {\cal A}_E/\kappa\approx 100-1000$,
indicating that the strong coupling regime is indeed 
feasible, the resolution of the proposed spectroscopy should be
very favorable to encourage experiments.

We acknowledge valuable
discussions with V.~Bouchiat, P.~Delsing, D.~Feinberg,
K.~Grove-Rasmussen, P.~Hyldgaard, G.~Johansson, H.~I.~J{\o}rgensen,
G.~Wendin, C.~Wilson, and A~Zazunov. The research was supported by
the Swedish Research Council (VR) and the Swedish Foundation for
Strategic Research (SSF), and the EC FP7 program
under grant agreement "SINGLE" No. 213609 .\\
\end{document}